\documentclass[aps, reprint, superscriptaddress, amsmath, amssymb, pra]{revtex4-2}

\usepackage[final]{graphicx}
\usepackage{amsmath}
\usepackage{amssymb}
\usepackage{float}
\usepackage[dvipsnames]{xcolor}
\usepackage[colorlinks=true,urlcolor=blue,linkcolor=blue,citecolor=blue, hypertexnames=false]{hyperref}
\usepackage{cleveref}
\usepackage{physics}
\usepackage[detect-weight=true,separate-uncertainty=true]{siunitx}
\usepackage[normalem]{ulem}

\crefname{equation}{Eq.}{Eqs.}
\Crefname{equation}{Equation}{Equations}
\crefname{figure}{Fig.}{Figs.}
\Crefname{figure}{Figure}{Figures}
\crefname{section}{Sec.}{Sects.}
\Crefname{section}{Section}{Sections}
\crefname{table}{Table}{Tables}
\crefname{appendix}{Appendix}{Apps.}
\Crefname{appendix}{Appendix}{Apps.}

\newcommand{\modf}[1]{{\color{blue}{#1}}}

\begin{document}

\title{Qubit readouts enabled by qubit cloaking}

\author{Manuel H. Mu\~{n}oz-Arias}
\email{munm2002@usherbrooke.ca}
\affiliation{Institut Quantique and Département de Physique, Université de Sherbrooke, Sherbrooke, Quebec, J1K 2R1, Canada}
\author{Crist\'obal Lled\'o}
\email{cristobal.lledo.veloso@usherbrooke.ca}
\affiliation{Institut Quantique and Département de Physique, Université de Sherbrooke, Sherbrooke, Quebec, J1K 2R1, Canada}
\author{Alexandre Blais}
\affiliation{Institut Quantique and Département de Physique, Université de Sherbrooke, Sherbrooke, Quebec, J1K 2R1, Canada}
\affiliation{Canadian Institute for Advanced Research, Toronto, M5G1M1 Ontario, Canada}

\date{\today}

\begin{abstract}
Time-dependent drives play a crucial role in quantum computing efforts with circuit quantum electrodynamics. They enable single-qubit control, entangling logical operations, as well as qubit readout. However, their presence can lead to deleterious effects such as large ac-Stark shifts and unwanted qubit transitions ultimately reflected into reduced control or readout fidelities. Qubit cloaking was introduced in~\textcite{QubitCloaking} to temporarily decouple the qubit from the coherent photon population of a driven cavity, allowing for the application of arbitrary displacements to the cavity field while avoiding the deleterious effects on the qubit. For qubit readout, cloaking permits to prearm the cavity with an, in principle, arbitrarily large number of photons, in anticipation to the qubit-state-dependent evolution of the cavity field, allowing for improved readout strategies. Here we take a closer look at two of them. First, \emph{arm-and-release} readout, introduced together with qubit cloaking, where after arming the cavity the cloaking mechanism is released and the cavity field evolves under the application of a constant drive amplitude. Second, an \emph{arm-and-longitudinal} readout scheme, where the cavity drive amplitude is slowly modulated after the release. We show that the two schemes complement each other, offering an improvement over the standard dispersive readout for any values of the dispersive interaction and cavity decay rate, as well as any target measurement integration time. Our results provide a recommendation for improving qubit readout without changes to the standard circuit QED architecture.
\end{abstract}

\maketitle


\section{Introduction}
\label{sec:intro}

Qubit readout is an indispensable operation in quantum information processing~\cite{NielsenChuangBook}. For superconducting qubits, the standard method is the dispersive readout which  consists in driving and measuring the response of a cavity whose resonant frequency is shifted depending on the state of a far detuned coupled qubit~\cite{Blais2004,Blais2021}. An advantage of dispersive readout is that at small measurement powers it is close to quantum non-demolition (QND)~\cite{Sank2016,Bultink2016,2019Touzard_PRL,QubitCloaking,Khezri2022}. However, non-QNDness at moderate power results in readout errors which have not yet reached the $10^{-3}$ level for measurement times of the order of $\SI{100}{ns}$~\cite{Walter2017, Sunada2022, Chen2023, Dassonneville2020}, lagging behind the best performance numbers of single-qubit and entangling gates~\cite{Negirneac2021,Ficheux2021,Xu2020,Ding2023}. Improving dispersive readout further is crucial to reach fault-tolerance in the circuit QED architecture for applications such as quantum error correction~\cite{Andersen2020,Krinner2022,Google2023,Zhao2022}.

In principle, increasing the strength of the cavity drive, leading to larger cavity photon population, can lead to improved readout~\cite{Blais2004}. However, this is not an ideal solution as even modest photon population can result in unwanted qubit transitions including leakage out of the transmon's computational subspace~\cite{Sank2016, Lescanne2019, Shillito2022, Cohen2023, Khezri2022}. Cavity drives also cause ac-Stark frequency shift of the qubit and broadening of the qubit linewidth via measurement-induced dephasing~\cite{Schuster2005,Boissonneault2009,Petrescu2020, Malekakhlagh2020}. Recently, an alternative approach to accelerating and improving the fidelity of dispersive readout has been demonstrated~\cite{QubitCloaking}. It is based on adding a cloaking drive on the qubit which allows the cavity to be armed with photons in a qubit-state-unconditional way. During the arming phase, the qubit is oblivious to the coherent state in the cavity thereby not experiencing ac-Stark shifts or measurement-induced dephasing. Once the desired mean photon population is reached, the cloaking mechanism can be released allowing the cavity field to evolve in a qubit-state-dependent way as in standard dispersive readout. In this paper, we explore how this additional knob---the armed photon population--- allows for optimizations of the cavity's phase-space  trajectories to maximize readout fidelity at short integration times.

One readout strategy where phase-space trajectories of the cavity field maximally distinguish the ground and excited states of the qubit at short measurement times is the longitudinal readout~\cite{Didier2015}. It relies on a qubit-resonator coupling between the qubit $\sigma^z$ and one of the resonator quadrature operators (e.g. the $P$ quadrature) with a modulated coupling frequency $g_z$. This interaction produces a cavity field displacement in opposite directions in phase space when the qubit is in the ground or excited state. While longitudinal interaction has been physically implemented in circuit QED~\cite{Roy2017,Eichler2018}, modulation of the coupling frequency has remained challenging, with no experimental demonstration so far of longitudinal readout except in its synthetic versions~\cite{2019Touzard_PRL, Ikonen2019}.

\begin{figure}[t!]
\centering{\includegraphics[width=0.85\linewidth]{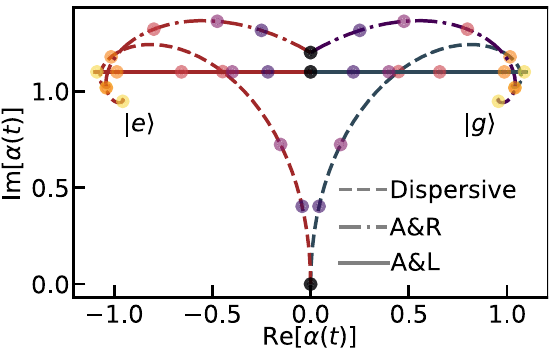}}
\caption{Path in phase-space
of the cavity amplitude when the qubit is in ground (blue) or excited (red) state for the three different schemes: dispersive (dashed line), arm-and-release (A\&R, dashed-dotted), and arm-and-longitudinal (A\&L, full line). The coloured dots indicate different times in the evolution, $\kappa t = 0,1,2,4,10,20$. The parameters are: $|\chi|/\kappa = 1$, and $\varepsilon_1/2\pi = \SI{19.85}{MHz}$ and $\SI{18.49}{MHz}$ as well as $\alpha_{\rm arm}/\sqrt{n_\text{max}} \approx 0.8$ and $1/\sqrt{2}$ for A\&R and A\&L, respectively.
}
\label{fig:figure_0}
\end{figure}

Building upon the arm-and-release idea, in this work we introduce a protocol for synthesizing a longitudinal readout process after the release step. The intuition is that the positions of the fixed point of the average cavity field---$\alpha_{g,e}^{\rm s}$ for the (g)round or (e)xcited qubit state---can be pushed or pulled by slowly modulating the drive amplitude, tailoring the trajectories followed by the time-dependent amplitudes $\alpha_{g,e}(t)$. This can be arranged in such a way that $\alpha_{g}(t)$ and $\alpha_{e}(t)$ separate from each other in exactly opposite directions in phase space, as under the longitudinal interaction, something we refer to as arm-and-longitudinal readout. This is illustrated in \cref{fig:figure_0} which shows the path in phase space of the cavity pointer states for the dispersive readout (dashed lines), the arm-and-release approach (dash-dotted lines), and the arm-and-longitudinal protocol (full lines). Crucially, this \textit{arm-and-longitudinal} readout protocol is only possible thanks to qubit cloaking~\cite{QubitCloaking}, as the cavity needs to start from an \textit{armed} state with non-zero field amplitude. To account for and avoid ionization~\cite{Sank2016, Lescanne2019, Shillito2022,Cohen2023}, it proves convenient that, under this modulation, the amplitudes $|\alpha_{g,e}(t)|$ increase monotonously in time until they reach a chosen maximum value, which can be taken to be below the ionization threshold. 

In this paper we show that either arm-and-release or arm-and-longitudinal always outperforms the standard dispersive readout for any measurement time and any ratio $\chi/\kappa$ of the dispersive interaction $\chi$ and the cavity decay rate $\kappa$. In particular, we show that for signal integration times $\gtrsim 10/\kappa$, the arm-and-longitudinal protocol outperforms the other two. These results can inform the design and manipulation of current and future superconducting qubit devices. Moreover, while we focus on circuit QED with superconducting qubits, these results are general and can be used in other platforms.

The rest of this manuscript is organized as follows. In \cref{sec:dispersive_ro} we briefly remind the reader about standard dispersive readout in the two-level system approximation, and in \cref{sec:cloaking} about qubit cloaking and the \textit{arm-and-release} scheme. In \cref{sec:longitudinal}, we explain how to transform arm-and-release into arm-and-longitudinal readout. Having introduced all the main ideas, in \cref{sec:comparison} we provide a performance comparison between the three schemes. In \cref{sec:full system dynamics} we show that the arm-and-longitudinal is valid beyond the two-level system approximation for the transmon, the rotating-wave approximation, and the dispersive approximation. Finally, in \cref{sec:outlook} we discuss our results and offer an outlook.


\section{Brief on dispersive qubit readout}
\label{sec:dispersive_ro}
We start by giving a short introduction to dispersive qubit readout in the two-level system approximation~\cite{Blais2004}. The Jaynes-Cummings Hamiltonian ($\hbar = 1)$
\begin{equation} \label{eq: JC Hamiltonian}
    \hat H_\mathrm{JC} = \omega_r \hat a^\dag \hat a + \frac{\omega_q}{2}\hat \sigma^z + g_c(\hat \sigma^+ \hat a  +  \hat \sigma^- \hat a^\dag)
\end{equation}
describes the interaction between a qubit and a cavity mode in the rotating-wave approximation. Here, $\hat a$ is the annihilation operator of the cavity mode of fundamental frequency $\omega_r$, $\hat \sigma^{z,+,-}$ are respectively the Pauli $z$, raising, and lowering operators for the qubit of fundamental frequency $\omega_q$, and $g_c$ is the interaction frequency. In the dispersive regime where $|\omega_q - \omega_r| \gg g_c$, the Hamiltonian is approximately diagonalized, up to second order in $g_c/|\omega_q - \omega_r|$, as
\begin{equation} \label{eq: dispersive Hamiltonian}
\hat U^\dag \hat H_{\rm JC} \hat U \approx \hat H_\mathrm{disp} =  \left(\omega_r + \frac{\chi}{2} \hat \sigma^z \right)\hat a^\dag \hat a + \frac{\tilde{\omega}_q}{2} \hat \sigma^z,
\end{equation}
where $\chi = 2g_c^2/(\omega_q-\omega_r)$ is the (full) dispersive interaction, $\tilde \omega_q = \omega_q + \chi/2$ is the Lamb-shifted qubit frequency, and $\hat{U}=\exp\{[g_c/(\omega_q-\omega_r)](\hat a^\dag \hat \sigma^- - \hat a \hat \sigma^+)\}$.

As the arrangement of the interaction term in \cref{eq: dispersive Hamiltonian} suggests, here we take the convention that the cavity frequency is pulled by $-\chi/2$ if the qubit is in the ground state $\ket{g}$ and by $+\chi/2$ if it is in the excited state $\ket{e}$. As a result, measuring the response of the cavity to an external drive, one can infer the state of the qubit.
For example, an initial condition $(c_g\ket{g} + c_e \ket{e})\ket{0}$ evolves to $c_g \ket{g,\alpha_g} + c_e \ket{e,\alpha_e}$ under the combined action of \cref{eq: dispersive Hamiltonian} and the drive 
\begin{equation}\label{eq: cavity drive}
\hat H_1(t) = -\varepsilon_1(t)(\hat a e^{i\omega_1 t} + \mathrm{h.c.})/2,    
\end{equation}
where $\ket{0}$ is the cavity vacuum state and $\alpha_{g,e}$ are coherent state amplitudes which act as pointer states and evolve as 
\begin{equation} \label{eq: readout semiclassical EOM}
\dot \alpha_{g,e} = -[i(\omega_r - \omega_1 \mp \chi/2 ) + \kappa/2]\alpha_{g,e} +i \varepsilon_1(t)/2
\end{equation}
in a frame rotating at the drive frequency $\omega_1$ and where $\varepsilon_1(t)$ is the cavity drive amplitude. \Cref{eq: readout semiclassical EOM} is obtained by using the dispersive approximation on 
the Lindblad master equation~\cite{Carmichael_lecture_notes, Breuer_book}
\begin{equation} \label{eq: Lindblad eq JC}
    \partial_t \hat \rho = -i[\hat H_\text{JC} + \hat H_1(t) , \hat \rho] + \kappa \mathcal D[\hat a]\hat \rho,
\end{equation}
where $\mathcal D[\hat a]\hat \rho \equiv \hat a \hat \rho \hat a^\dag - (1/2)\{\hat a^\dag \hat a , \hat \rho\}$. For simplicity, one assumes in \cref{eq: readout semiclassical EOM} that the qubit state remains constant.


\section{Qubit cloaking and arm-and-release readout}
\label{sec:cloaking}

As introduced in Ref.~\cite{QubitCloaking}, cloaking a qubit in a cavity allows to decouple the qubit from the classical part of a cavity field. This approach, which is achieved with an appropriately chosen drive on the qubit, has been shown to lead to faster dispersive readout via an arm-and-release (A\&R) scheme. The latter consists in arming the cavity with photons while cloaking the qubit, and to subsequently release the cloaking mechanism allowing the dispersive readout dynamics to proceed. Noting from \cref{eq: readout semiclassical EOM} that the path of the coherent states $\alpha_{g,e}$ will (for $\omega_1 =\omega_r$) initially separate from one another along the $X$ quadrature at a speed $\mathrm{Re}[\dot \alpha_g - \dot\alpha_e] \propto \chi \sqrt{\bar n(t)}$, where $\bar n(t)$ is the mean photon number, the A\&R approach has the clear advantage of having a finite initial speed ($\bar n(0)\neq 0)$ thanks to the armed photons, unlike in dispersive readout where $\bar n(0)=0$. Moreover, since high-fidelity qubit gates can be realized on a cloaked qubit, the arming time does not factor into the readout time~\cite{QubitCloaking}.

The intuition behind the cloaking mechanism is the following. The cavity drive \cref{eq: cavity drive} results in a coherent state inside the cavity which acts as an effective classical drive on the qubit~\cite{Cohen2023}. An additional drive on the qubit can destructively interfere with this effective drive, leaving the qubit to experience the cavity as if it was in the vacuum state. To make this observation more precise, consider applying a displacement transformation $\hat a \to \hat a + \alpha(t)$ on \cref{eq: Lindblad eq JC} with $\alpha(t)$ chosen to eliminate the effect of the cavity drive. Due to the qubit-cavity coupling term $(\propto g_c)$, this transformation effectively passes the drive to the qubit and the displaced Hamiltonian reads $\hat H_{\rm JC} + g_c(\alpha(t) \hat \sigma^+ + \alpha^*(t)\hat \sigma^-)$. With an additional qubit cancellation drive
\begin{equation}
\hat H_2(t) = -g_c(\alpha(t) \hat \sigma^+ + \alpha^*(t) \hat \sigma^-),
\end{equation}
the effective drive on the qubit is cancelled out in the displaced frame, where the Hamiltonian is just the undriven $\hat H_{\rm JC}$ of \cref{eq: JC Hamiltonian}. Here, the complex amplitude $\alpha(t)$ corresponds to the coherent state amplitude of the cavity field in the absence of the qubit, namely, it evolves according to $\dot \alpha(t) = -(i\omega_r + \kappa/2) \alpha(t) + i\varepsilon_1(t)e^{-i\omega_1 t}/2$ in the laboratory frame. The above results are exact and can be made to account for the rapidly rotating terms which were (implicitly) ignored in \cref{eq: JC Hamiltonian}, as well as account for the multilevel nature of superconducting qubits. Moreover, the result remains unchanged if qubit dissipation and dephasing are included~\cite{QubitCloaking}. The picture emerging from qubit cloaking is that the cavity field is displaced in phase space by an amount $\alpha(t)$ that is qubit-state independent. Equivalently, the qubit is not affected by this displacement.

Having introduced the fundamentals of qubit cloaking and described the steps involved in arm-and-release readout, we study the expected performance of this approach. Since the arming time does not factor into the readout time, for the rest of this section we assume the intracavity field to start at a chosen position along the P quadrature in phase space. That is, without loss of generality we take a purely imaginary $\alpha(0) = i\alpha_{\rm arm}$ with some real and positive $\alpha_{\rm arm}$. Furthermore, during the release step ($t>0$) we take the amplitude $\varepsilon_1$ of the cavity drive to remain constant. With these two conditions, we can integrate \cref{eq: readout semiclassical EOM} to obtain
\begin{equation} \label{eq:amplitude_arm_and_release}
\begin{split}
&\alpha_{e}(t) = \alpha_{\rm arm}\left[ \sin(\frac{\chi t}{2}) +i\cos(\frac{\chi t}{2}) \right]e^{-\frac{\kappa}{2}t}  \\ 
+& \tilde{\varepsilon}_1\left\{\chi - \left[\chi\cos(\frac{\chi t}{2}) + \kappa\sin(\frac{\chi t}{2})\right]e^{-\frac{\kappa}{2}t} \right\}  \\
+& i\tilde{\varepsilon}_1\left\{\kappa + \left[\chi\sin(\frac{\chi t}{2}) - \kappa\cos(\frac{\chi t}{2})\right]e^{-\frac{\kappa}{2}t} \right\},
\end{split}
\end{equation}
where $\tilde{\varepsilon}_1 = \varepsilon_1/(\chi^2 + \kappa^2)$ and the expression for $\alpha_g(t)$ is obtained by replacing $\chi\to -\chi$. We recover the result expected for the standard dispersive readout when taking $\alpha_{\rm arm} \to 0$ in \cref{eq:amplitude_arm_and_release}. By noting that the expressions for $\alpha_e(t)$ and $\alpha_g(t)$ differ only in their real part, the contribution of the term proportional to $\alpha_\text{arm}$ to the measurement signal is $\propto \text{Re}[\alpha_g(t) - \alpha_e(t)] = 2\alpha_{\rm arm}\sin(|\chi|t/2)e^{-\kappa t/2}$. At short times this contribution is positive, enriching the signal.

For fixed parameters, the amplitudes $\alpha_{g,e}$ reach the same steady states for both dispersive and A\&R readout, but the maximum mean photon number $\bar n_{\rm max}$ visited during the trajectories are not the same---see~\cref{app:stuff_arm_and_release} for an explicit expression for $\bar{n}(t)$. Since qubit ionization sets a maximum mean photon number for quantum nondemolition readout~\cite{Shillito2022}, to make a fair comparison we first choose $(\chi, \kappa, \varepsilon_1)$ for the standard dispersive dynamics and then we adjust $\varepsilon_1$ for A\&R such that both schemes share the same $\bar{n}_\text{max}$. In \cref{fig:figure_0} we compare the phase-space trajectories for dispersive (dashed lines) and arm-and-release (dashed-dotted lines). While A\&R is clearly faster than dispersive (see the colored dots), fixing the maximum photon number makes the two fixed points of A\&R have a smaller separation than those of dispersive. As we will discuss in \cref{sec:comparison}, there is an important trade-off between speed at short times and long-time state discrimination that influence the performance of the different readout schemes.

\begin{figure}[t!]
\includegraphics[width=0.85\linewidth]{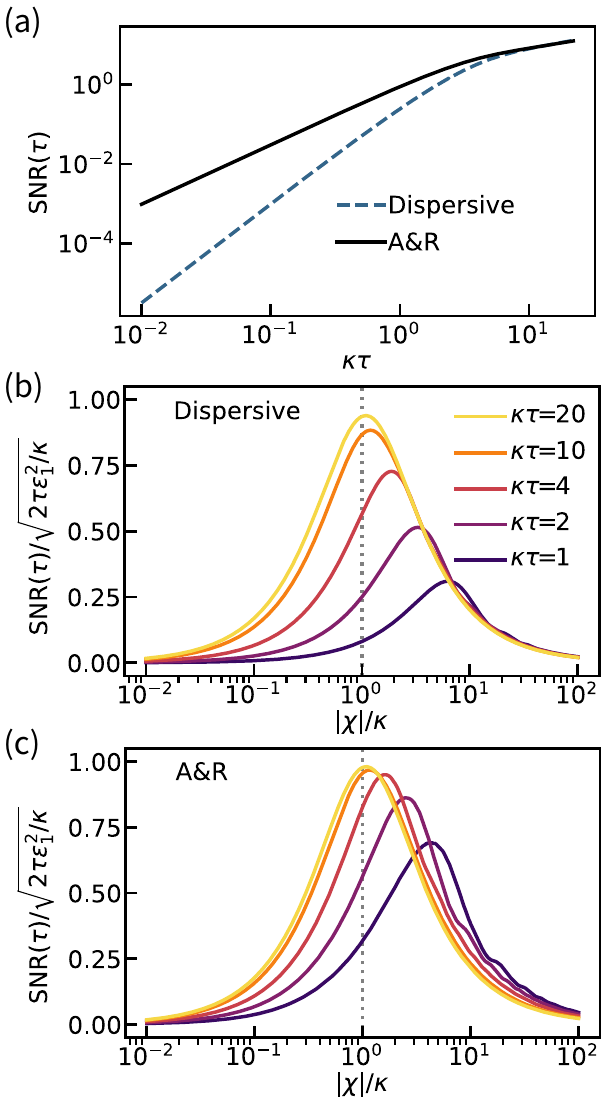}
\caption{(a) Signal-to-noise ratio (SNR) as a function of measurement integration time for dispersive (dashed) and A\&R (solid) readouts, for the corresponding pairs of trajectories shown in \cref{fig:figure_0}. (c,d) Normalized SNR vs $|\chi|/\kappa$ for dispersive (b) and A\&R (c) readout. The different curves corresponds to different measurement integration times in units of $1/\kappa$. }
\label{fig:figure_1}
\end{figure}

To compare the performance of the A\&R approach to dispersive readout, we compute for both cases the signal-to-noise ratio (SNR) for homodyne detection of the $X$-quadrature of the cavity output field. The corresponding integrated measurement operator up to time $\tau$ is given by $\hat{M}(\tau) = \sqrt{\kappa}\int_0^\tau dt K(t)[\hat{a}_{\rm out}^\dagger(t) + \hat{a}_{\rm out}(t)]$~\cite{Bultink2018,Blais2021}. In this expression, $K(t)$ is a filter function and $\hat{a}_{\rm out}$ represents the output field related to the intracavity field via the input-output boundary condition $\hat{a}_{\rm out} = \hat{a}_{\rm in} + \sqrt{\kappa}\hat{a}$, with $\hat{a}_{\rm in}$ the input field~\cite{GardinerBook}. The signal is defined as $|\langle \hat{M}(\tau) \rangle_{e} - \langle\hat{M}(\tau)\rangle_g|$ and the imprecision noise is given by the fluctuations around the mean-field of $\hat{M}$, that is, $\hat{M}_N(\tau) = \hat{M}(\tau) - \langle\hat M(\tau)\rangle$. Putting these two elements together, the squared SNR can be expressed as~\cite{Blais2021}
\begin{equation} \label{eq: signal_to_noise}
{\rm SNR}^2 = \frac{|\langle \hat{M}(\tau) \rangle_{e} - \langle\hat{M}(\tau)\rangle_g|^2}{\langle\hat{M}^2_N(\tau)\rangle_e + \langle\hat{M}^2_N(\tau)\rangle_g}.
\end{equation}
The filter $K(t)$ should favorably weight times where the signal is larger, thus naturally the optimal filter is $K(t) = |\langle \hat M (t) \rangle_e - \langle \hat M(t) \rangle_g|$~\cite{Gambetta2007, Bultink2018}. Assuming that the input and output fields remain coherent and in the limit of unit-efficiency measurement one obtains~\cite{Bultink2018}
\begin{equation} \label{eq: snr_optimal_filter}
{\rm SNR}^2 = 2\kappa \int_0^\tau dt\, \text{Re}[\alpha_e(t) - \alpha_g(t)]^2.
\end{equation}
Even though the full analytical expression of the SNR for A\&R is rather long and thus uninformative (see~\cref{app:stuff_arm_and_release}), there are some interesting properties worth discussing. As pointed out previously, the A\&R scheme receives a boost compared to dispersive readout stemming from the nonzero photons in the initial condition. This is manifested in the SNR whose short-time scaling $(\kappa\tau\ll1)$ reads
\begin{equation} \label{eq: snr_arm_release_short}
{\rm SNR}_{\rm A\&R} \approx \sqrt{\frac{2}{3}}  \frac{\alpha_{\rm arm}|\chi|}{\kappa}(\kappa\tau)^{3/2},
\end{equation}
and is to be contrasted with the short-time scaling of standard dispersive readout
\begin{equation} \label{eq: snr_disp_short}
{\rm SNR}_{\rm disp} \approx \frac{1}{8}\sqrt\frac{3}{2}
\frac{\varepsilon_1 |\chi|}{\kappa^2}(\kappa\tau)^{5/2}.
\end{equation}
With A\&R we thus acquire signal at a faster rate proportional to $\alpha_{\rm arm}$. Remarkably, this rate is equal to the rate exhibited by the ideal longitudinal qubit readout~\cite{Didier2015}. We illustrate this in \cref{fig:figure_1}b by comparing the SNR as a function of the integration time for the case of $|\chi|/\kappa = 1$ which is optimal for the dispersive readout~\footnote{\modf{We note that for fixed drive amplitude, large $|\chi|/\kappa$ ratio, and long measurement integration times, the optimal driving frequency for dispersive readout becomes $\omega_1 = \omega_r \pm (1/2)\sqrt{\chi^2 - \kappa^2}$. If, however, we fix the maximum number of photons during readout, $\omega_1 =\omega_r$ is always the optimal driving frequency and the optimal value of $|\chi|/\kappa$ tends towards one as the integration time grows.}}. In the short measurement time regime, the different scalings are evident from the slopes of the dashed-dotted line (arm-and-release) and the dashed line (dispersive), furthermore, even at $\kappa\tau \to0$ a clear improvement is observed, corroborating our observation of the enhancement provided by a nonzero $\alpha_{\rm arm}$.

In the opposite limit of long measurement times, $\kappa\tau\gg 1$, the SNR is dominated by the term 
\begin{equation} \label{eq: snr_arm_release_long}
{\rm SNR}_{\rm A\&R} \approx \sqrt{8}\frac{\varepsilon_1}{\kappa}\frac{|\chi|/\kappa}{1 + ({\chi}/{\kappa})^2} \sqrt{\kappa\tau},    
\end{equation}
which is the same long-measurement-time SNR as for dispersive readout. This is because at long times the contribution coming from $\alpha_\text{arm}$ has decayed, as is evident from the first line of \cref{eq:amplitude_arm_and_release}. As such, in this asymptotic limit the ratio $|\chi|/\kappa=1$ corresponds to the optimal working point for both A\&R and dispersive readout. To illustrate this,  \cref{fig:figure_1}b,c shows the SNR at different measurement times versus $|\chi|/\kappa$ for dispersive (b) and A\&R (c) readouts, respectively. In both cases, the maximum approaches $\chi/\kappa = 1$ at long times. It is also worth noticing that for short measurement times the optimal value of $|\chi|/\kappa$ is $>1$ in both cases, with A\&R reaching the asymptotic value faster. Moreover, for each of the measurement times used in \cref{fig:figure_1}b,c, A\&R has a larger maximum SNR than dispersive as a consequence of the boost coming from $\alpha_{\rm arm}$.


\section{Arm-and-longitudinal readout}
\label{sec:longitudinal}

As anticipated in the introduction, arm-and-release (A\&R) readout can be transformed into arm-and-longitudinal (A\&L) readout by the appropriate choice of the time-dependent drive amplitude $\varepsilon_1(t)$. Let us start with 
a simple intuition behind this scheme. As shown in \cref{fig:figure_0}, the A\&R amplitudes $\alpha_{g,e}(t)$ curve towards the real phase-space axis as time increases, moving towards their fixed points $\alpha^{\rm s}_{g,e} = \varepsilon_1/(\mp\chi - i\kappa)$. To make the coherent amplitudes separate along the $X$-quadrature in a straight line without curving, we slowly increase $\varepsilon_1(t)$ in time so as to push away the fixed points the trajectories follow.

The explicit form of $\varepsilon_1(t)$ needed to obtain longitudinal-like displacements of the cavity field $\alpha_{g,e}(t)$ can be obtained from \cref{eq: readout semiclassical EOM} by imposing the displacement to be only along the $X$ quadrature ($\omega_{\rm r} = \omega_1$). With the initial condition $\alpha(0) = i\alpha_{\rm arm}$, we obtain (see \cref{app:integral_equation_amplitude})
\begin{equation} \label{eq: eps_sol}
\varepsilon_1(t) = \alpha_{\rm arm} \frac{\chi^2}{\kappa}\left( 1 - e^{-\frac{\kappa}{2}t} \right) + \alpha_{\rm arm}\kappa.
\end{equation}
Using this expression, the qubit-state dependent coherent amplitudes take the desired longitudinal-like motion along the real phase-space axis
\begin{equation} \label{eq: alpha_long}
\alpha_{g,e}(t) = \mp\alpha_{\rm arm}\frac{\chi}{\kappa}\left(1 - e^{-\frac{\kappa}{2}t} \right) + i\alpha_{\rm arm}.
\end{equation}
This arm-and-longitudinal scheme requires $\alpha_{\rm arm}\ne0$ and is thus enabled by our ability to prearm the cavity field using qubit cloaking~\cite{QubitCloaking}. We remark that the frequency of the drive needs to be chosen in between the two qubit-state-dependent frequency responses of the cavity ($\omega_r\pm \chi/2$), which is important to ensure that the drive modulation necessary for arm-and-longitudinal is independent of the qubit state.

In \cref{fig:figure_0} we show the A\&L trajectories in phase space corresponding to \cref{eq: alpha_long} (solid lines), comparing them to the A\&R and the dispersive readouts. For a fair comparison, the three pairs of trajectories visit the same maximum mean photon number.
While at short times the coherent amplitudes $\alpha_{g,e}(t)$ separate faster for A\&R, at long times the separation is larger for A\&L, hinting at an interesting trade-off between the two schemes. We can anticipate that, depending on the ratio $|\chi|/\kappa$ and the aimed measurement integration time, one or the other might be the best strategy. We devote \cref{sec:comparison} to this analysis.

Replacing \cref{eq: alpha_long} into \cref{eq: snr_optimal_filter}, we find for the SNR of the A\&L readout
\begin{equation} \label{eq: snr_long}
{\rm SNR}_{\rm A\&L} = 
\sqrt{8}\frac{\alpha_{\rm arm}|\chi|}{\kappa}
\sqrt{\kappa \tau - 3 + 4e^{-\frac{\kappa}{2}\tau}  - e^{-\kappa\tau}}.
\end{equation}
At short times $\kappa \tau \ll 1$, it takes the simpler form
\begin{equation}\label{eq: snr_long_short}
{\rm SNR}_\text{A\&L} \approx \sqrt{\frac{2}{3}}\frac{\alpha_{\rm arm}|\chi|}{\kappa}(\kappa\tau)^{3/2},
\end{equation}
and thus offers better performance than the dispersive readout whose SNR is $\propto(\kappa\tau)^{5/2}$, see \cref{eq: snr_disp_short}.
We note that the expression in \cref{eq: snr_long} has the same functional form as the SNR obtained for a modulated longitudinal qubit-cavity interaction~\cite{Didier2015}, whose Hamiltonian is
\begin{equation}\label{eq: Hz}
  H_z(t) = ig_z(t)\hat{\sigma}^z(\hat{a}^\dagger - \hat{a}), 
\end{equation}
with $\tilde g_z$ the amplitude of $g_z(t)$ which is modulated at the cavity frequency, provided that one includes the optimal filter~\footnote{Following Ref.~\cite{Didier2015} under a longitudinal interaction the cavity amplitude evolves as $\alpha_{g,e}(t) = \mp\frac{\tilde{g}_z}{\kappa}\left( 1 - e^{-\kappa t/2}\right)$, thus we can readily integrate \cref{eq: snr_optimal_filter} and obtain ${\rm SNR}^2 = 8\frac{\tilde{g}_z^2}{\kappa^2}\left( \kappa\tau - 3 + 4e^{-\kappa\tau/2} - e^{-\kappa\tau} \right)$}
and identifies $\alpha_{\rm arm}|\chi|$ with $\tilde g_z$.

This comparison suggests that, in a prearmed cavity, the product $\alpha_{\rm arm}|\chi|$ plays the role of an effective interaction driving the readout dynamics. Importantly, realizing the Hamiltonian \cref{eq: Hz} requires a qubit-cavity coupling that is different from the standard capacitive coupling of circuit QED which rather leads to \cref{eq: JC Hamiltonian}~\cite{Didier2015,Risher2016,Billangeon2015}. In contrast, our A\&L approach can be implemented without changes to the standard circuit QED architecture~\cite{Blais2021}.

Although A\&R and A\&L share the same short-time SNR scaling, the values of the armed amplitude, $\alpha_{\rm arm}$ in \cref{eq: snr_arm_release_short,eq: snr_long_short}, are not the same.
For fixed ratio of $|\chi|/\kappa$, measurement time and maximum mean photon number $\bar n_\text{max}$, one can optimize over $\alpha_{\rm arm}$ to maximize the SNR of arm-and-release. On the other hand, for arm-and-longitudinal readout the value of the initial amplitude $\alpha_{\rm arm}$ is fixed and from \cref{eq: alpha_long} it reads
\begin{equation} \label{eq: alpha_arm_long}
\alpha_{\rm arm} = \sqrt{\frac{\bar n_{\rm max}}{1 + \left({\chi}/{\kappa}\right)^2}}.    
\end{equation}
The reason is just that for A\&L the mean photon number reaches its maximum in the steady state, compare the full and dashed-dotted lines in \cref{fig:figure_0}.

\begin{figure*}[t!]
\includegraphics{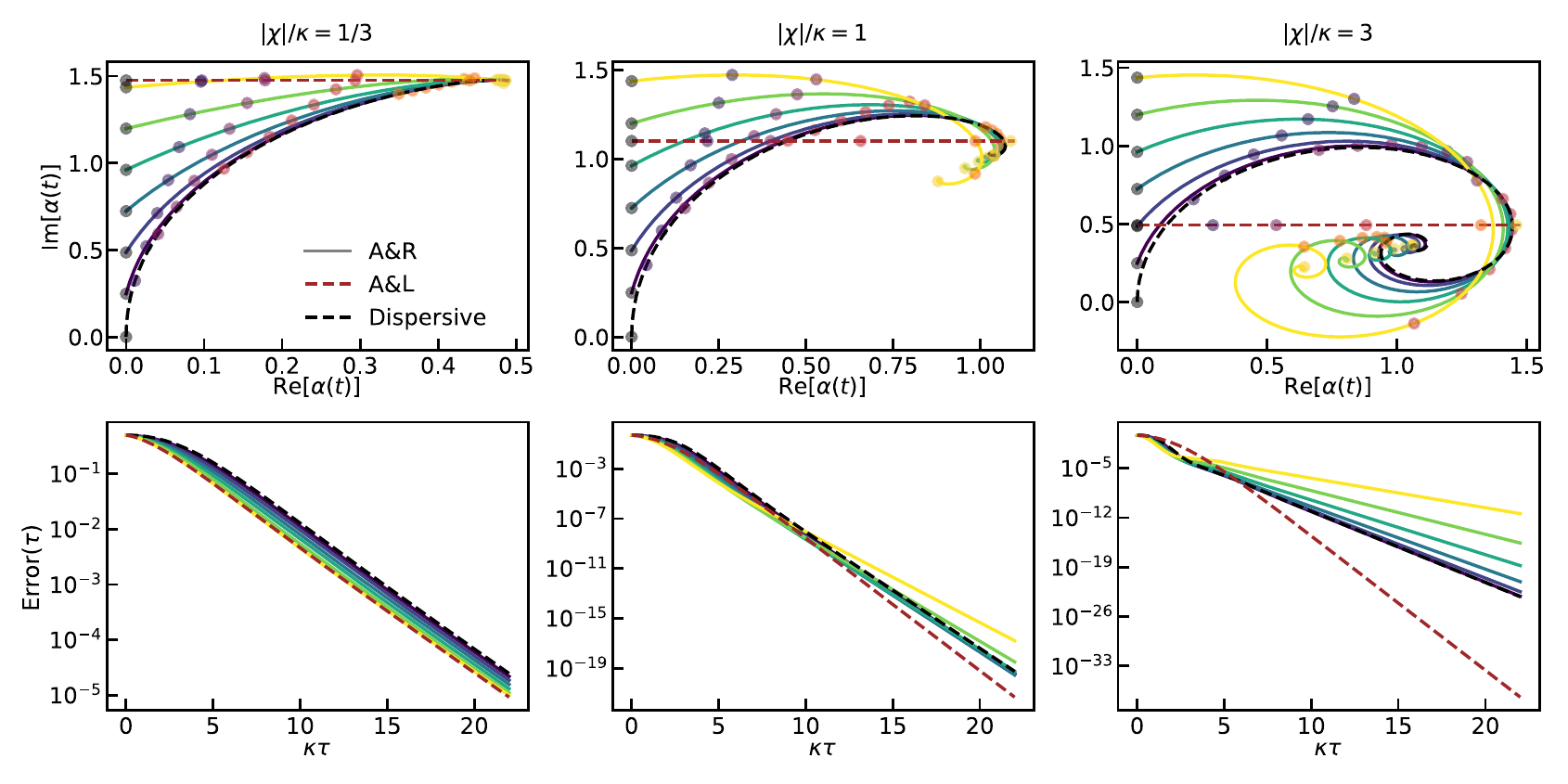}
\caption{Top row: Path in phase-space of the cavity amplitude $\alpha_g(t)$ for the three different schemes: standard dispersive (black dashed line), arm-and-release (A\&R, full lines), and arm-and-longitudinal (A\&L, red dashed line). The different colours for A\&R trajectories indicate different initial amplitudes $\alpha_\text{arm}$, while the coloured dots indicate different times in the evolution, $\kappa t = 0,1,2,4,10,20$. All trajectories, in all three panels, visit the same maximum mean photon number $n_\text{max} = 2.44$. Bottom row: Assignment error as function of measurement integration time for the corresponding trajectories depicted in the top row's panels. The parameters are from left to right: $|\chi|/\kappa = 1/3$, $1$, $3$, and, for A\&L, $\alpha_{\rm arm}/\sqrt{n_\text{max}} = \sqrt{9}/\sqrt{10}$, $1/\sqrt{2}$, $1/\sqrt{10}$. For dispersive, $\varepsilon_1/2\pi = \SI{15.77}{MHz}$, $\SI{19.85}{MHz}$, $\SI{34.38}{MHz}$. For A\&R, $\alpha_{\rm arm}\in\left(0, \sqrt{n_{\rm max}} \right]$ and $\varepsilon_1$ is obtained, after fixing $\alpha_\text{arm}$, from the constraint of having less than the maximum mean photon number.}
\label{fig:figure_2}
\end{figure*}

For long times, $\kappa\tau \gg 1$, we obtain 
\begin{equation} \label{eq: snr_long_long}
\begin{split}
{\rm SNR}_\text{A\&L} \approx& \sqrt{8}\frac{\alpha_{\rm arm}|\chi|}{\kappa} \sqrt{\kappa\tau} \\
&= \sqrt{8 n_{\rm max}}\frac{|\chi|/\kappa}{\sqrt{1 + \chi^2/\kappa^2}}\sqrt{\kappa\tau},
o\end{split}
\end{equation}
where in the second line we have used \cref{eq: alpha_arm_long}. This has the same long-time scaling with measurement time as dispersive and arm-and-release readout.  Fixing $\bar n_{\rm max}$ and the integration time, the SNR saturates at large $|\chi|/\kappa$.


\section{Comparison of the readout approaches}
\label{sec:comparison}

Let us now put in perspective the results and discussions of the previous three sections. We have introduced an extra level of time-dependent control to extend arm-and-release to arm-and-longitudinal readout. Both of these schemes have a short-time SNR which scales with time as the ideal longitudinal readout~\cite{Didier2015}, and which offers an improvement over that of dispersive readout. For the opposite limit of long times, all three schemes have the same scaling with time. These two observations beg the questions of which one is the superior scheme and how much do we gain with it. We will see that this question does not have a simple answer, and in fact, depending on the device parameters represented by the ratio $|\chi|/\kappa$ and the measurement time (see the examples given in \cref{sec:intro}), either arm-and-release or arm-and-longitudinal is the best alternative.  

Let us start this discussion with a close inspection of phase-space trajectories of the three readout schemes. In top row of \cref{fig:figure_2} we show the trajectories for $\alpha_g(t)$---since $\alpha_e(t)$ is its mirror image---for the ratios $|\chi|/\kappa = 1/3, 1, 3$. In all three panels, the full lines show the trajectories for arm-and-release for some illustrative values of $\alpha_{\rm arm}$, the black dashed line shows the trajectory of dispersive readout, and the red dashed line of arm-and-longitudinal. Same color dots on different trajectories represent the positions of the evolving amplitude at equal times ($\kappa t = 0, 1, 2, 4, 10, 20$). Importantly, across all three panels and all trajectories the maximum mean photon number that is attained is the same.

As can be observed, interestingly when $|\chi|/\kappa$ is small $\alpha_{\rm arm}$ in arm-and-release can be used as a control parameter to interpolate between dispersive ($\alpha_{\rm arm} \to 0$) and arm-and-longitudinal ($\alpha_{\rm arm} \to \sqrt{n_{\rm max}})$. Notice, however, that for arm-and-release the constraint of a fixed maximum mean photon number implies that the drive amplitude depends on the value of the initial cavity amplitude, i.e., $\varepsilon_1 \to \varepsilon_1(\alpha_{\rm arm})$. To see this, consider a fixed value of the ratio $|\chi|/\kappa$. Then, as the value of $\alpha_{\rm arm}$ is increased, the maximum mean photon number is reached at shorter and shorter times. As a consequence, the value of $\varepsilon_1$ decreases, thus effectively moving the steady-state fixed points $\alpha^{\rm s}_g$ closer to the origin. This can be seen more markedly in the top row of \cref{fig:figure_2} for $|\chi|/\kappa = 3$, indicating the existence of a trade-off between the speed of evolution at short times and the separation distance between the steady-state fixed points. As a result, for large values of $|\chi|/\kappa$, arm-and-release might not provide an advantage over dispersive readout, an observation which can be made precise by studying the SNR. Indeed, in the intermediate to long-time regime the SNR is dominated by a term proportional to $\varepsilon_1$, c.f. \cref{eq: snr_arm_release_long}, thus having a $\varepsilon_1$ which decreases with increasing $\alpha_{\rm arm}$ highly diminishes the performance of the scheme. 

In contrast, the horizontal trajectory of arm-and-longitudinal avoids altogether this trade-off. In fact, we are guaranteed to reach the maximum separation distance between the time-evolving amplitudes at steady state. As such, arm-and-longitudinal might be slower at short times than, say arm-and-release for some values of $\alpha_{\rm arm}$, as can be appreciated from the equal time dots in \cref{fig:figure_2} top row. However, in the regime of intermediate to long times,   the position of the time-evolving amplitude is further away from the $P$-quadrature phase-space axis as the ratio $\chi/\kappa$ is increased than in the other two readout schemes. This yields a purely geometrical advantage for arm-and-longitudinal over the other two approaches.

It is also useful to consider the assignment (or single-shot) measurement error defined as $E_m = (1/2)[P(e|g) + P(g|e)]$, where $P(n|m)$ is the probability of assigning the qubit to be in the state $n$ when it was actually in $m$~\cite{Magesan2015}. For the Gaussian distributions we are considering, the assignment error is related to the SNR~\cite{Blais2021} by $E_m = (1/2){\rm erfc}({\rm SNR}/2)$, where ${\rm erfc}$ is the complementary error function. In the bottom row of \cref{fig:figure_2} we plot the assignment error for each of the trajectories of the top row in the same figure. For $|\chi|/\kappa=1/3$ in \cref{fig:figure_2}, the error of the A\&R scheme (solid lines) always improves over the error of dispersive readout (black dashed line), and at larger $\alpha_{\rm arm}$ it is always better. Upon increasing the value of $|\chi|/\kappa$ there is crossing of the curves corresponding to the different schemes. For short times, the largest $\alpha_{\rm arm}$ results in the smallest error, yet for larger times its performance is the worst. This is in agreement with the geometric picture of the trajectories, which is most evident in panel for $|\chi|/\kappa=3$ where the largest arming amplitude gives the worst performance and instead A\&L readout gives a large improvement (note the change in the range of the vertical axis amongst the three panels). This observation opens the door to achieve improved readout performance in parameter regimes far from the well known optimal working point $|\chi|/\kappa =1$ of dispersive readout.

\begin{figure*}[t!]
\includegraphics{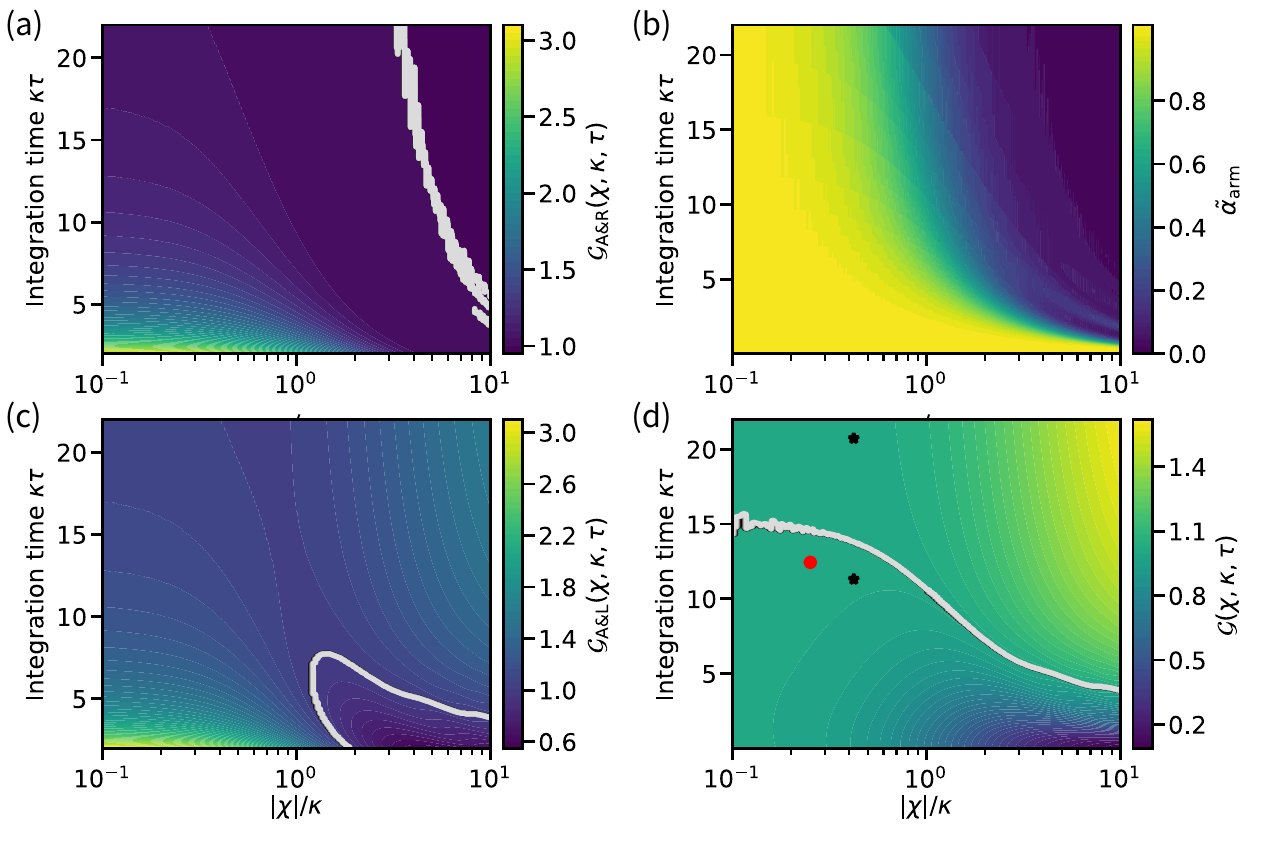}
\caption{(a) SNR relative gain between A\&R and dispersive, \cref{eq: relative_gain_arm_release}, as a function of $|\chi|/\kappa$ and measurement integration time. (b) Optimal value of the (normalized) amplitude $\tilde \alpha_\text{arm}$ associated to the best SNR in panel (a). (c) SNR relative gain between arm-and-longitudinal and dispersive, \cref{eq: relative_gain_long}, as a function of $|\chi|/\kappa$ and measurement integration time. (d) Ratio between the relative gains of A\&L and A\&R, \cref{eq: absolute_gain}. The black stars correspond to the parameters of the experiment in Ref.~\cite{Walter2017} and the red dot to those of Ref.~\cite{QubitCloaking} (in which A\&R was used), see text. In (a,c,d) the gray contours indicate the separation between the regions with different optimal readout strategy. In all panels, for a given $|\chi|/\kappa$ all other parameters are fixed by the choice of $n_{\rm max}$.}
\label{fig:figure_3}
\end{figure*}

\subsection{Arm-and-release readout vs dispersive readout}
Up to this point we have presented qualitative and quantitative arguments in support of an improvement offered by arm-and-release readout over dispersive readout, noticing that there may exists a region in the parameter space $(|\chi|/\kappa, \kappa\tau)$ where dispersive readout is the superior scheme. In this subsection, we quantify the extent of the improvement offered by arm-and-release as well as identify the region in parameter space where this improvement is guaranteed.

To quantify the improvement offered by arm-and-release readout, we introduce the relative gain
\begin{equation} \label{eq: relative_gain_arm_release}
\mathcal{G}_{\rm A\&R}( \chi,\kappa,\tau) = \frac{\underset{\alpha_{\rm arm}}{\max}\left\{\rm SNR_{\rm A\&R}(\chi, \kappa, \tau, \alpha_{\rm arm})\right\} }{{\rm SNR_{\rm disp}(\chi, \kappa, \tau)}},
\end{equation}
where ${\rm SNR_{\rm disp}(\chi, \kappa, \tau)} = {\rm SNR_{\rm A\&R}(\chi, \kappa, \tau, \alpha_{\rm arm} = 0)}$. The maximization is taken considering the constraint that both methods should lead to the same maximum mean photon number. In \cref{fig:figure_3}a we show this relative gain as a function of the ratio $|\chi|/\kappa$ and the measurement time $\kappa\tau$. We observe a large region of this parameter space where arm-and-release is advantageous, reaching gains as large as $\mathcal{G}_{\rm A\&R} \sim 400$ for $\kappa \tau \ll 1$ (not shown) and small $|\chi|/\kappa$, and $\mathcal{G}_{\rm A\&R} \sim 2$ at intermediate times. However, this advantage decreases at large $|\chi|/\kappa$ and for readout times ranging from intermediate, $\kappa\tau \sim 4.5$, all the way to the asymptotic limit where dispersive readout is advantageous. The gray region in \cref{fig:figure_3}a delimits the transition between the two regimes where one or the other scheme performs better. This is consistent with the geometrical argument for the trajectories discussed before, by which at intermediate to long times the coherent amplitudes are in close proximity to the steady-state fixed points, and these are closer to the origin for A\&R when $|\chi|/\kappa$ is large.

In \cref{fig:figure_3}b we show the normalized value of the amplitude, $\tilde \alpha_\text{arm} = \alpha_\text{arm}^\text{optimal}/\sqrt{\bar n_\text{max}}$, that maximizes the relative gain $\mathcal{G}_\text{A\&R}$ of panel (a). The normalization bounds it to $\tilde \alpha_\text{arm} \in [0,1]$, making it universal. Hence, the optimal value of $\alpha_\text{arm}$ for any system parameters can be deduced from \cref{fig:figure_3}b by rescaling $\tilde \alpha_\text{arm}$ by the appropriate maximum mean photon number that is desired. The reason behind this is that all relative gains remain unchanged under the rescaling $\alpha_\text{arm} \to \alpha_\text{arm}/ \sqrt{\bar n}$ and $\varepsilon_1 \to \varepsilon_1 / \sqrt{\bar n}$, as it constitutes shrinking the whole phase space to the region at the interior of the unit circle. This panel gives an interpretation to the region on the right of the grey contour in \cref{fig:figure_3}a, where A\&R does not offer an advantage over dispersive. Since in this region $\tilde{\alpha}_{\rm arm}\sim1$ the maximum mean photon number of the phase-space paths occurs at times $\kappa\tau\rightarrow0$ thus leading to fixed points $\alpha_{g,e}^s$ which are almost at the origin.

\subsection{Arm-and-longitudinal readout vs dispersive readout}

While there is a region of parameter space where dispersive readout is favourable with respect to A\&R, c.f. \cref{fig:figure_3}a, we now show that A\&L offers a large improvement over A\&R in that region. To quantify this, we introduce the relative gain of A\&L as 
\begin{equation} \label{eq: relative_gain_long}
\mathcal{G}_{\rm A\&L}(\chi,\kappa, \tau) = \frac{{\rm SNR}_{\rm A\&L}(\chi, \kappa, \tau, \alpha_{\rm arm})}{{\rm SNR_{\rm disp}(\chi, \kappa, \tau)}}.
\end{equation}
In \cref{fig:figure_3}c we show this relative gain in the parameter space $(|\chi|/\kappa, \kappa\tau)$. The A\&L readout scheme offers a large gain $\mathcal{G}_{\rm A\&L}\sim400$ at $\kappa\tau\ll 1$ (not shown), similar to the relative gain of arm-and-release, a consequence of the finite armed photons. Overall, arm-and-longitudinal performs better than dispersive, except in the region of large $|\chi|/\kappa$ and short measurement times, which is delimited by the gray contour in \cref{fig:figure_3}c. In particular, in the region 
where the advantage of A\&R over dispersive was not guaranteed, now A\&L is the best of the three methods (compare \cref{fig:figure_3}a and c). In short, qubit cloaking always allows to improve readout fidelity regardless of the value of $|\chi|/\kappa$ and the measurement integration time.

We stress that even a modest gain $\mathcal{G}_{\rm A\&L}\sim 2$ will yield a large improvement for the measurement discrimination error. This is illustrated in \cref{fig:figure_2} which shows the assignment error $E_m$ for  $|\chi|/\kappa = 3$ where A\&L gives an improvement of several orders of magnitude at intermediate to long times. In fact, for a large SNR, $E_m = (1/2)\text{erfc}(\text{SNR}/2)\approx e^{-\text{SNR}^2/4}/(\sqrt{\pi} \text{SNR})$, which means that a larger prefactor in the long-time scaling of the SNR can have a huge impact in the reduction of the error.

\subsection{Arm-and-release vs. arm-and-longitudinal: which one should you use?}

The results of the previous two subsections are summarized in \cref{fig:figure_3}d, where we show the ratio
\begin{equation} \label{eq: absolute_gain}
\begin{split}
\mathcal{G}(\chi, \kappa, \tau) =& \frac{\mathcal{G}_{\rm A\&L}(\chi, \kappa, \tau)}{\mathcal{G}_{\rm A\&R}(\chi, \kappa \tau)} \\
=& \frac{{\rm SNR}_{\rm A\&L}(\chi, \kappa, \tau, \alpha_{\rm arm})}{\underset{\alpha_{\rm arm}}{\max}\left\{\rm SNR_{\rm A\&R}(\chi, \kappa, \tau, \alpha_{\rm arm})\right\} }
\end{split}
\end{equation}
between the SNRs of A\&L and A\&R. Together these two schemes always provide a better strategy than standard dispersive readout on current superconducting circuit experiments. All that is left is thus for us to make the recommendation of how, given the device parameters $|\chi|/\kappa$ and a target measurement integration time, to make the more out of this pair of schemes. The gray contour  in \cref{fig:figure_3}d indicates $\mathcal{G}(\chi,\kappa,\tau) = 1$, separating the region where A\&L or A\&R is more advantageous. In the case where A\&R is the preferable choice, one can resort to \cref{fig:figure_3}b to identify the appropriate arming amplitude yielding the maximum improvement.

As an example, the experiment of Ref.~\cite{Walter2017} performed readout of a transmon qubit with $|\chi|/\kappa = 0.42$ in $\kappa\tau = 11.31$ and $20.73$ with $98.25$\% and $99.2$\% average fidelity, respectively. In \cref{fig:figure_3}d we show these two configurations (black stars), with our recommendation being A\&R and A\&L, respectively. We also indicate with a red dot the parameters of the \emph{qubit cloaking} experiment in which A\&R was used~\cite{QubitCloaking}. In all cases the expected gain on SNR over dispersive readout is in the 20--30\% range. Albeit modest, it can signify a large improvement in the discrimination error and thus the readout fidelity, as mentioned in the discussion of \cref{fig:figure_2} bottom row. Depending on the value of the SNR, the error can be reduced by up to 8\%.


\section{Full system dynamics}
\label{sec:full system dynamics}

\begin{figure}[t!]
\includegraphics{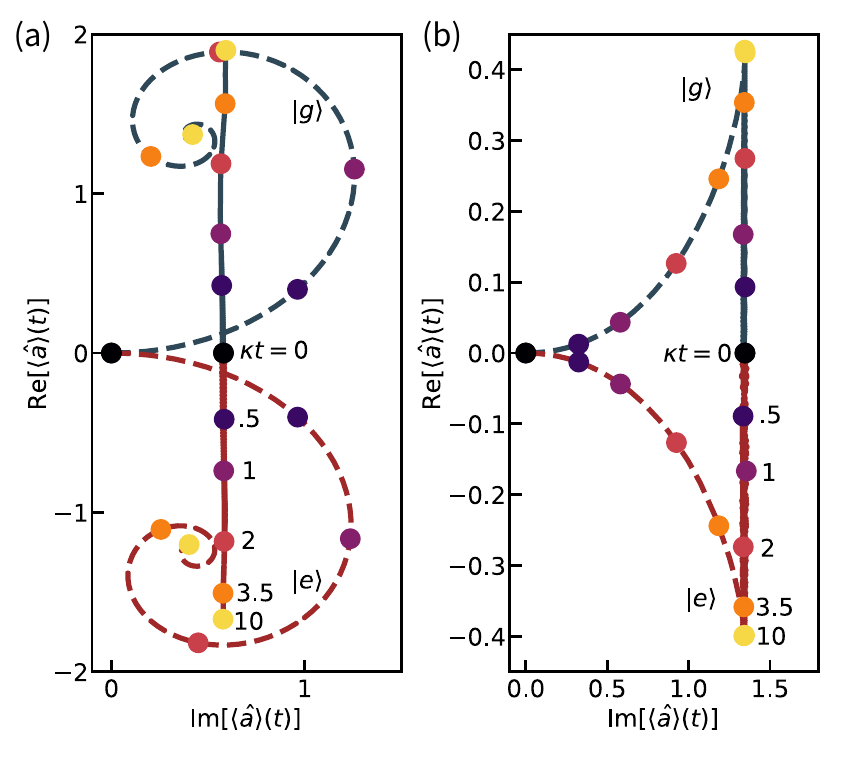}
\caption{Numerically obtained paths of the cavity pointer states in phase space using a transmon multilevel Hamiltonian, no RWA, and no dispersive approximation for A\&L (full lines) and standard dispersive (dashed lines). In (a) [(b)] we use $\kappa/2\pi = \SI{1}{MHz}$ $(\SI{10.1}{MHz})$, and for A\&L readout we choose $n_\mathrm{tar}=2$ ($1$), which is in between zero photons and the maximum number attained $\bar{n}_\mathrm{max}=4$ ($2$). This results in $|\chi_{n_\mathrm{tar}}|/\kappa =3.286$ in (a) and 0.326 in (b). For the standard dispersive case, we use $\omega_1/2\pi = \SI{7.665}{GHz}$ ($\SI{7.666}{GHz}$) and $\varepsilon_1/2\pi = \SI{4.876}{MHz}$ ($\SI{15}{MHz}$) for (a) [(b)]. All other system parameters read $E_J/2\pi = \SI{16.93}{GHz}$, $E_C/2\pi = \SI{200.4}{MHz}$, $g/2\pi = \SI{159.1}{MHz}$, and $\omega_r/2\pi = \SI{7.655}{GHz}$.
}
\label{fig:figure_full_transmon}
\end{figure}

To simplify the presentation, the above discussion relied on the following three approximations: (i) truncating the multilevel nonlinear system to a two-level system, (ii) the rotating-wave approximation, and (iii) the dispersive approximation. We now show that the readout trajectories can be made longitudinal-like even without using these approximations.  As a concrete example, we consider a transmon qubit~\cite{Koch2007} coupled capacitevely to a cavity with the Hamiltonian ($\hbar =1$)
\begin{equation} \label{eq: Full transmon and cavity Hamiltonian}
    \hat H_0 = \omega_r \hat a^\dag \hat a + 4 E_C \hat n_{\rm tr}^2 - E_J \cos(\hat \varphi_{\rm{tr}}) + ig\hat n_{\rm tr} (\hat a^\dag - \hat a).
\end{equation}
In this expression, $\hat n_{\rm tr}$ and $\hat \varphi_{\rm tr}$ are the transmon charge and phase operators, and $E_C$, $E_J$, and $g$ the charging, Josephson, and coupling frequencies, respectively~\cite{Blais2021}. To this Hamiltonian we add the cavity drive term $\hat H_1 (t)= i\varepsilon_1(t)\sin(\omega_1 t)(\hat a^\dag - \hat a)$, and solve numerically the Lindblad equation 
\begin{equation}
    \partial_t \hat \rho = -i[\hat H_0 + \hat H_1(t), \hat \rho] + \kappa \mathcal D[\hat a] (\hat \rho).
\end{equation}
We modulate $\varepsilon_1(t)$ as in~\cref{eq: eps_sol}. Given that the arming time does not factor into the readout time, we begin our simulation for A\&L, shown in \cref{fig:figure_full_transmon}b, with a preloaded cavity, and simulate the readout dynamics following the release of the cloaking mechanism.

In \cref{fig:figure_full_transmon} we show the numerically obtained readout phase-space trajectories for A\&L (full lines) and the standard dispersive (dashed lines) for two values of the ratio $|\chi_{n_\mathrm{tar}}|/\kappa \approx 3.3$ (a) and $|\chi_{n_\mathrm{tar}}|/\kappa \approx 0.3$ (b). 
Here $\chi_{n_\mathrm{tar}}$ is the numerically obtained dispersive shift evaluated at a target photon number $n_\mathrm{tar}$, see \cref{app:numerical_chi} for details.
The colored dots corresponds to times $t=0$, $0.5$, $1$, $2$, $3.5$ and $10$ in units of $1/\kappa$. This figure shows that, even when using the full transmon Hamiltonian and not using the rotating-wave nor dispersive approximations, the readout trajectories can be made longitudinal-like via the modulation of the drive amplitude as in \cref{eq: eps_sol}. Moreover, in agreement with Figs.~\ref{fig:figure_2} and~\ref{fig:figure_3}, the best improvement of A\&L over standard dispersive is obtained for long times when $|\chi|/\kappa>1$, as is evident here from the much larger separation of the average cavity amplitudes for A\&L readout in panel (a) [notice the change in vertical scale between the two panels]. The only system parameter that we change in (a) and (b) is $\kappa/2\pi$, taking values $\SI{1}{MHz}$ and $\SI{10.1}{MHz}$, respectively. We note that the asymmetry observed between the ground and excited state paths in phase space is due to Purcell decay which affects both dispersive and A\&L in similar ways. As discussed in Ref.~\cite{QubitCloaking}, qubit cloaking and thus A\&L works with minimal change in the presence of a Purcell filter.


\section{Discussion and outlook}
\label{sec:outlook}
We have studied the arm-and-release and arm-and-longitudinal readout schemes which are derive from qubit cloaking~\cite{QubitCloaking}. Both are a direct consequence of our ability to prearm the cavity with photons in anticipation to the readout dynamics. These two approaches complement each other and together offer a modest advantage over the standard dispersive readout for all values of the ratio $|\chi|/\kappa$ and measurement times. Since qubit cloaking relies on standard circuit QED hardware~\cite{QubitCloaking}, this advantage can be achieved free of hardware overhead, that is, no alteration to this architecture is required. Arm-and-longitudinal is realized by a slow turn-off of the cavity readout drive. The resulting control over the cavity mean photon population makes this scheme suitable to prevent undesirable measurement-induced transitions and qubit ionization. We thus hope that our results will help current and future circuit QED experiments to perform readout at or close to the state-of-the-art level while relaxing some parameter optimizations.

An interesting future avenue of research which might yield improved readout schemes is to exploit the complementarity of the two schemes discussed in this work. The combination of rapid evolution under arm-and-release for short times, $\kappa\tau \sim 2$, followed by time modulation of the drive amplitude could bring together the best of both schemes. This hybrid strategy is guaranteed to succeed provided $\text{Im}[\alpha_g(t_0)] = \text{Im}[\alpha_e(t_0)]$ is treated as a new initial condition for the dynamical problem [c.f.~the discussion around~\cref{eq: eps_sol}]. 

The longitudinal phase-space evolution of the cavity amplitude during arm-and-longitudinal may also be useful beyond the task of qubit readout. For instance the error-correction cycles with the GKP code use phase-space motions of the cavity amplitude which are quasi-longitudinal~\cite{Eickbusch2022,Sivak2023}, it is interesting to explore to what extent this error correction schemes might see an improvement using the protocol presented in this work. Furthermore, the longitudinal phase-space motion could also be useful in implementing cavity-assisted two-qubits gates that rely on the geometric phase of the cavity field~\cite{Paik2016}.


\section*{Acknowledgements}
The authors are grateful to Audrey Bienfait, Benjamin Huard and Rémy Dassonneville for helpful discussions and comments on the manuscript. This work is supported by a collaboration between the US DOE and other Agencies. This material is based upon work supported by the U.S. Department of Energy, Office of Science, National Quantum Information Science Research Centers, Quantum Systems Accelerator. Additional support is acknowledged from NSERC, the Canada First Research Excellence Fund, and the Minist\`ere de l’\'Economie et de l’Innovation du Qu\'ebec.


\bibliography{references}

\appendix


\section{Mean photon number and signal-to-noise ratio for arm-and-release readout}
\label{app:stuff_arm_and_release}
In this appendix we present the expressions for the time dependence of the mean photon number and the signal-to-noise ratio for the arm-and-release scheme, as well as for standard dispersive 
readout which is recovered in the limit $\alpha_\text{arm}\to 0$. The mean photon number is directly computed from \cref{eq:amplitude_arm_and_release} as $\bar{n}(t) = |\alpha(t)|^2$, and we obtain
\\
\begin{equation} \label{eq: photon_arm_and_release}
\begin{split}
&\bar{n}_{\rm A\&R}(t) = \frac{\varepsilon_1^2}{\chi^2 + \kappa^2}(1 + e^{-\frac{\kappa}{2} t}) + \alpha_\text{arm}^2 e^{-\kappa t}  \\
&+ \frac{\varepsilon_1}{\chi^2 + \kappa^2}\bigg[2\alpha_\text{arm} \left[\chi \sin(\frac{\chi}{2}t) + \kappa \cos(\frac{\chi}{2}t) - \kappa e^{-\frac{\kappa}{2} t} \right]  \\
&\qquad \qquad \qquad - 2\varepsilon_1 \cos(\frac{\chi}{2}t) \bigg]e^{-\kappa t/2}
\end{split}
\end{equation}
\\
The signal-to-noise ratio is computed directly from \cref{eq: snr_optimal_filter}. We find
\begin{widetext}
\begin{eqnarray} \label{eq: snr_arm_and_release}
&&\frac{\text{SNR}_\text{A\&R}^2}{8\kappa} = \frac{\varepsilon_1^2 \chi^2}{(\kappa^2 + \chi^2)^2} \tau - \frac{4\varepsilon_1^2\chi^2}{(\chi^2 + \kappa^2)^3}\left[ \kappa - \left( \kappa\cos(\frac{\chi}{2}\tau) - \chi \sin(\frac{\chi}{2}\tau) \right)e^{-\frac{\kappa}{2}\tau} \right] \nonumber \\
&+& \frac{\varepsilon_1^2\chi^2}{2\kappa(\chi^2 + \kappa^2)^3} \left[ \chi^2 + 2\kappa^2 - \left( \chi^2 + \kappa^2 + \kappa^2 \cos(\chi\tau) - \chi\kappa \sin(\chi\tau) \right)e^{-\kappa\tau} \right] \nonumber \\
&-&\frac{4\varepsilon_1^2\kappa\chi}{(\chi^2 + \kappa^2)^3} \left[ \chi - \left( \cos\left(\frac{\chi}{2}\tau \right) + \kappa \sin\left(\frac{\chi}{2} \tau\right) \right)e^{-\frac{\kappa}{2}\tau}\right] + \frac{\varepsilon_1^2\kappa\chi}{(\chi^2 + \kappa^2)^3}\left[ \chi - ( \chi\cos(\chi\tau) + \kappa\sin(\chi\tau))e^{-\kappa\tau} \right] \nonumber \\
&+& \frac{\varepsilon_1^2\kappa}{2(\chi^2 + \kappa^2)^3} \left[ \chi^2 - (\chi^2 + \kappa^2 - \kappa^2\cos(\chi\tau) + \kappa\chi\sin(\chi\tau))e^{-\kappa\tau} \right] \nonumber \\
&+& \frac{4\alpha_{\rm arm}\varepsilon_1\chi}{(\chi^2 + \kappa^2)^2} \left[ \chi - \left( \chi\cos\left(\frac{\chi}{2}\tau\right) + \kappa\sin\left(\frac{\chi}{2}\tau \right) \right)e^{-\frac{\kappa}{2}\tau} \right] - \frac{\alpha_{\rm arm}\varepsilon_1\chi}{(\chi^2 + \kappa^2)^2} \left[ \chi - \left( \chi\cos(\chi\tau) + \kappa\sin(\chi\tau) \right)e^{-\kappa\tau}  \right] \nonumber \\ 
&+& \frac{\alpha_{\rm arm}^2}{2\kappa(\chi^2 + \kappa^2)}\left[ \chi^2 - \left( \chi^2 + \kappa^2 - \kappa^2 \cos(\chi\tau) + \kappa\chi\sin(\chi\tau) \right)e^{-\kappa\tau} \right] \nonumber \\ 
&-& \frac{\alpha_{\rm arm}\varepsilon_1}{(\chi^2 + \kappa^2)^2} \left[ \chi^2 - \left( \chi^2 + \kappa^2 - \kappa^2\cos(\chi\tau) + \kappa\chi\sin(\chi\tau) \right)e^{-\kappa\tau} \right]
\end{eqnarray}
\end{widetext}


\section{Explicit form of the drive amplitude for arm-and-longitudinal readout}
\label{app:integral_equation_amplitude}
As mentioned in \cref{sec:longitudinal}, our starting point is the equation of motion for the intracavity field amplitude with a resonant drive, $\omega_{1} = \omega_r$, given by 
\begin{equation} \label{eq: eom_appendix}
\dot \alpha_{g,e} = -(\mp i\chi/2 + \kappa/2)\alpha_{g,e} +i \varepsilon_1(t)/2,
\end{equation}
and its formal solution, 
\begin{equation} \label{eq: formal_sol_EOM}
\begin{split}
\alpha_e(t) =& \alpha(0)e^{-\left(i\frac{\chi}{2} + \frac{\kappa}{2}\right)t} \\
&+ \frac{i}{2}\int_0^t e^{-\left(i\frac{\chi}{2} + \frac{\kappa}{2}\right)(t - t')}\epsilon(t')dt',
\end{split}
\end{equation} 
with $\alpha_g(t) = \alpha_e(t; \chi \to -\chi)$.
Furthermore, given that our initial condition is ${\rm Re}[\alpha(0)] = 0$ and ${\rm Im}[\alpha(0)] = \alpha_{\rm arm}$ with $\alpha_{\rm arm}$ a positive real number, we achieve our goal of making the trajectories separate along the $X$-quadrature if the condition ${\rm Im}[\alpha_{g,e}(t)] = \alpha_{\rm arm}$ is satisfied for all times $t\geq0$. This immediately implies $\frac{d}{dt}{\rm Im}[\alpha(t)]=0$, and after plugging this into \cref{eq: eom_appendix} we obtain
\begin{equation} \label{eq: eps_equation1}
\epsilon_1(t) = \chi {\rm Re}[\alpha(t)] + \kappa {\rm Im}[\alpha(t)],
\end{equation}
where we dropped the sign dependence on $\chi$, as this will not affect the final answer which is qubit-state independent.

At this point, we split \cref{eq: formal_sol_EOM} into its real and imaginary parts, and plug them into \cref{eq: eps_equation1} to obtain 
\begin{equation} \label{eq: eps_equation2}
\begin{split}
\varepsilon_1(t) = 2\alpha_\text{arm} K(t,0) + \int\limits_0^t d\tau K(t,\tau)\varepsilon_1(\tau),
\end{split}
\end{equation}
where the kernel reads
\begin{equation} \label{eq: kernel}
\begin{split}
K(t, \tau) =& \frac{1}{2}\left[\chi \sin(\frac{\chi}{2} (t - \tau)) + \kappa \cos(\frac{\chi}{2} (t - \tau)) \right] \\
\qquad \quad& \times e^{-\frac{\kappa}{2}(t - \tau)},
\end{split}
\end{equation}
The expression in Eq.~(\ref{eq: eps_equation2}) is a Volterra integral equation of the second kind (see chapter 16 of Ref.~\cite{Arfken1967}). Importantly, this kernel is separable, meaning that it can be written as
\begin{equation}
K(t, \tau) = \sum_{l=1}^L G_l(t)W_l(\tau),
\end{equation}
for some integer $L$ and some functions $\{G_l\}$ and $\{W_l\}$. This fact is a sufficient condition for the integral equation to have a unique solution (see chapter 16 of~Ref.\cite{Arfken1967}). This solution reads 
\begin{equation}
\epsilon_1(t) = \frac{\alpha_{\rm arm}\chi^2}{\kappa}\left( 1 - e^{-\frac{\kappa}{2}t} \right) + \alpha_{\rm arm}\kappa,
\end{equation}
given in \cref{eq: eps_sol} in the main text.

\section{Numerical determination of $\chi$ for the full-cosine transmon model}
\label{app:numerical_chi}
With the approximations introduced in \cref{sec:dispersive_ro}, the dispersive interaction takes the usual form $\approx \chi \hat \sigma^+\hat \sigma^- \hat a^\dag \hat a$, where the dispersive shift $\chi$ is independent of the photon number. Here for a multilevel system, using perturbation theory to approximately diagonalize the system Hamiltonian gives
\begin{equation} \label{eq: dispersive approximation full Hamiltonian}
\hat U^\dag \hat H_0 \hat U \approx \tilde \omega_r \hat a^\dag \hat a + \sum_i \tilde \epsilon_i \ket{i}\bra{i} + \sum_{i,n} n \chi_{i,n} \ket{i,n}\bra{i,n},
\end{equation}
where $\tilde \omega_r$ is the Lamb-shifted cavity frequency and the dispersive interaction $\chi_{i,n}$ depends now on the cavity Fock number $n$~\cite{Blais2021}. Here, $\{\tilde \epsilon_i\}$ and $\{\ket{i}\}$  with $i=g,e,f,\dots$ correspond respectively to the Lamb-shifted energies and the eigenstates of the transmon Hamiltonian $4E_C \hat n_\mathrm{tr} - E_J \cos(\hat \varphi_\mathrm{tr})$. The actual eigenstates of $\hat H_0$, labeled $\ket{\overline{i,n}} = \hat U \ket{i,n}$ have associated eigenvalues $\epsilon_{\overline{i,n}} \approx (\tilde \omega_r + \chi_{i,n})n + \tilde \epsilon_i$ according to \cref{eq: dispersive approximation full Hamiltonian}. We can define energy branches for each of the transmon states, see Ref.~\cite{Shillito2022}. The branch associated with the transmon being in the ground (excited) state corresponds approximately to cavity energies $(\tilde \omega_r + \chi_{g(e),n})n$. Choosing an optimal drive frequency for readout with a target Fock number $n_\mathrm{tar}$ corresponds to using $\omega_1 = \tilde \omega_r + (\chi_{g,n_\mathrm{tar}} + \chi_{e,n_\mathrm{tar}})/2$, exactly in between the two branches at that Fock number. The full dispersive shift at $n_\mathrm{tar}$ is given by $\chi_{n_\mathrm{tar}}\equiv \chi_{e,n_\mathrm{tar}} - \chi_{g,n_\mathrm{tar}}$. In the modulation of the drive amplitude, \cref{eq: eps_sol}, instead of $\chi$ we use $\chi_{n_\mathrm{tar}}$ with a chosen $n_\mathrm{tar}$. Our protocol works perfectly well as long as $\chi_n$ does not change dramatically from the initial $n\sim \bar n = |\alpha_{\rm arm}|^2$ mean photon number to the steady-state $n \sim \bar{n}^{\rm s} = n_\text{max}$ maximum mean photon number. In particular, the dynamics should not lead to photon numbers so large that transitions to higher transmon energy levels ($i=f, \dots$) are induced~\cite{Sank2016, Lescanne2019, Shillito2022, Cohen2023, Khezri2022}.

\end{document}